\documentclass[a4paper]{IEEEtran}

\usepackage[utf8]{inputenc}
\usepackage{csquotes}
\usepackage{graphicx}
\usepackage{amssymb}
\usepackage{amsmath}
\usepackage{amsthm}
\usepackage{eufrak}
\usepackage{mathtools}
\usepackage{dot2texi}
\usepackage{tikz}
\usepackage{algorithm}
\usepackage{algpseudocode}
\usepackage{listings}
\usepackage{wrapfig}
\usepackage{subcaption}
\usepackage{multicol}
\usepackage{enumitem}

\usepackage[english]{babel}

\usepackage{xcolor}
\usepackage[breaklinks]{hyperref}

\usepackage{booktabs}

\usepackage{cleveref}

\newcommand{\cn}[1]{\textsuperscript{[citation needed]}}

\newcommand{\intgsym}{\mathcal{I}}
\newcommand{\confsym}{\mathcal{C}}

\newcommand{\intg}[1]{{#1}_{\intgsym}}
\newcommand{\conf}[1]{{#1}_{\confsym}}
\newcommand{\invar}[1]{\mathit{#1}}
\newcommand{\outvar}[1]{\overrightarrow{\mathit{#1}}}

\newcommand{\PR}{\mathrm{P}}

\newcommand{\R}{\mathrm{R}}

\DeclareMathOperator{\true}{\mathit{True}}
\DeclareMathOperator{\false}{\mathit{False}}

\lstdefinestyle{XML} {
    language=XML,
    extendedchars=true, 
    breaklines=true,
    breakatwhitespace=true,
    emph={},
    emphstyle=\color{red},
    basicstyle=\ttfamily,
    columns=fullflexible,
    commentstyle=\color{gray}\upshape,
    morestring=[b]",
    morecomment=[s]{<?}{?>},
    morecomment=[s][\color{green}]{<!--}{-->},
    keywordstyle=\color{blue},
    stringstyle=\ttfamily\color{purple}\normalfont,
    tagstyle=\color{blue}\bf,
    morekeywords={attribute,xmlns,version,type,release},
}

\newcommand{\encryptrule}{\intg{\outvar{Ciphertext}} \rightarrow{} \intg{\invar{Plaintext}}\label{eq:encrypt_1}\\ &\land{}\intg{\invar{Key}}\land{}\conf{\invar{Key}}\label{eq:encrypt_2}\\ &\land{}\intg{\invar{Ctr}}\label{eq:encrypt_3}}

\newcommand{\projecttitle}{\emph{PrettyCat}}

\title{\projecttitle: Adaptive guarantee-controlled software partitioning of security protocols}
\author{Alexander Senier \and Martin Beck \and Thorsten Strufe\\\{alexander.senier\textbar martin.beck1\textbar thorsten.strufe\}@tu-dresden.de}

\hypersetup
{
    colorlinks  = true,
    linkcolor   = blue,
    citecolor   = blue ,
    pdfauthor   = {Alexander Senier, Martin Beck, Thorsten Strufe},
    pdfkeywords = {application partitioning, security protocols, trusted computing base}
}

\begin{document}

\maketitle

\begin{abstract}

One single error can result in a total compromise of all security in today's
large, monolithic software. Partitioning of software can help simplify code-review
and verification, whereas isolated execution of software-components limits the
impact of incorrect implementations.

However, existing application partitioning techniques are too expensive, too imprecise,
or involve unsafe manual steps. An automatic, yet safe, approach to dissect
security protocols into component-based systems is not available.

We present a method and toolset to automatically segregate security related software
into an indefinite number of partitions, based on the security guarantees
required by the deployed cryptographic building blocks. As partitioning imposes communication overhead,
we offer a range of sound performance optimizations.

Furthermore, by applying our approach to the
secure messaging protocol OTR, we demonstrate its applicability and achieve a
significant reduction of the trusted computing base.
Compared to a monolithic implementation, only 29\% of the partitioned protocol
requires confidentiality guarantees with a process overhead comparable to
common sandboxing techniques.\\[1em]


\end{abstract}

\section{\label{sec:intro}Introduction}

Mass surveillance is reality today. It is technically possible to record the
Internet traffic of a whole country~\cite{devereaux2014data} or monitor
complete Internet exchange points in real time~\cite{meister2015how}.
Consequently, software vendors realized that encryption is key to protect the
sensitive information of their customers. Strong security protocols gradually
became a
default~\cite{schillace2010default,twitter2012securing,koum2016end},
some of the protocols even got formally
verified~\cite{bhargavan2013implementing,cohn2016formal,kobeissi2017automated},
which means that it can only be broken if an assumption is invalidated, or a
mathematically hard problem is solved.

In practice, security protocols are always embedded into a very complex
software system. Depending on their realization, they rely on operating system
services, application runtimes or web browsers. If an attacker exploits a
vulnerability in any of these dependencies, security protocols can be broken or
bypassed, even when formally verified.

To achieve correctness, we must ensure an error-free implementation of the
security protocol and all its dependencies. However, the size of those software
dependencies typically is in the range of some million lines of code. This, by
far, exceeds the limits of formal verification and makes thorough manual review
infeasible.

To enable complete verification, the fact that only small parts of a security
protocol implementation are security critical can be leveraged. If software is
segregated into multiple isolated components which interact only through
well-defined interfaces, verification can be done on a per-component basis.
While the small critical components are analyzed independently with manageable
effort, large uncritical components can be ignored completely.

Component-based architectures realize this concept using for example a microkernel. This
small software isolates components and allows access to resources and
communication channels only if permitted explicitly. This default-deny policy
and the limited functionality of the microkernel dramatically reduced the
trusted computing base (TCB), i.e. the code that has to be correct to fulfill
the objective of a security protocol.

While the component-based approach has been studied extensively in the past,
its application to security protocol implementations poses a number of open
problems: (a) Determining security critical and uncritical components
systematically with minimum manual intervention and identifying data-flows
between them is complex and error-prone. (b)
Minimizing the TCBs of the resulting component-based system is unsolved.
(c) There are no methods that guarantee the overall security of componentized
implementations. (d) No systematic approach for reducing the introduced overhead
is available.

To solve these problems, we model security protocol implementations as a
graph of connected primitives annotated with predicates representing the
required security guarantees. A constraint data flow analysis assigns valid
guarantees to the model using an SMT solver. In a subsequent partitioning step,
primitives with compatible guarantees are combined into components to optimize
performance. The security-performance trade-off made during partitioning is
configurable.

\subsection*{Our contribution}

To the best of our knowledge, we are the first to present an automatic,
yet sound partitioning algorithm that can minimize the required guarantees and can be tweaked
regarding its security-performance trade-off. Our main contributions in the field
of software partitioning are:
\begin{itemize}
  \item a methodology for predicate-based data-flow analysis for assigning security guarantees
  \item combination of benefits from data-flow analysis and manual decomposition
  \item trade-off between security and performance using optimization criteria.
\end{itemize}

Further results, like the component-based, minimal-TCB implementation of the OTR messaging protocol,
as well as the combination of predicate logic and software-partitioning together with
a set of predicate templates for common security-primitives is of independent interest.


\Cref{sec:related_work} discusses related work and distinguishes our method from existing approaches.
Subsequently we present our approach, introduce necessary building blocks and demonstrate it's applicability using two well-known examples in \Cref{sec:design}.
We scale our method to larger protocols using our automated toolkit described in \Cref{sec:scaling}.
\Cref{sec:discussion} puts our results into context regarding the related work and discusses future research directions.
A summary concludes our work in \Cref{sec:summary}.

\section{\label{sec:related_work}Related work}
Our analysis relates to many well established research areas. We classify previous results into {\em a) \textbf{protocol proofs}} for cryptographic protocols based on presumably hard mathematical problem, {\em b) \textbf{software verification}} to show the correctness of an implementation, {\em c) \textbf{decomposition mechanisms}} for partitioning software into multiple components, and {\em d) \textbf{isolation mechanisms}} for securely executing mutually distrusted processes. In the following paragraphs we position the proposed analysis along these classes, clarify our contribution, highlight important results over previous works and remove ambiguity.

\paragraph{\textbf{Protocol proofs}}

Some cryptographic protocols can be proven to be secure as long as a hard mathematical problems cannot be solved and the assumptions are not invalidated. As such, the well known textbook Diffie-Hellman key agreement protocol is secure as long as the discrete logarithm problem can not be solved easily, or the assumption that the adversary can not intercept and modify messages between both participating parties is invalidated. Relying on proven cryptographic protocols guarantees that the cryptographic building blocks are not easily attackable. Security does, however, still depend on a correct software implementation of the proven protocol. We use these results to systematically eliminate attacks on the theoretical basis of secure systems.

\paragraph{\textbf{Software verification}}

Architecing secure software systems requires the correct implementation of software components. To find programming mistakes and software errors, methods for software verification were developed. Applying these to software implementations guarantees the absence of runtime errors, otherwise potentially nullifying the proven security properties.
A formally verified reference implementation of the Transport Layer Security
(TLS) protocol is presented in~\cite{bhargavan2013implementing}. The authors
show confidentiality and integrity of data sent over the protocol.

Domain specific languages can be used for writing cryptographic protocol code that can be checked afterwards. \emph{ProScript}~\cite{KobeissiBhargavanBlanchetEuroSP17} is such an example that can be executed within JavaScript
programs and used for symbolic analysis. The authors verify a variant of the
Signal Protocol for secure messaging.

Even though the software implementing a cryptographic protocol is checked, errors are likely to be found in the surrounding runtimes, libraries and
operating systems required to run the verified cryptographic protocol code. Verified implementations may serve as a correct input model for our approach, but do not solve the problem to find small trusted computing bases.

\paragraph{\textbf{Decomposition mechanisms}}

As verified software for proven cryptographic protocols is still vulnerable just by its huge trusted computing bases, isolated execution of decomposed software was proposed. There exist two main approaches for software decomposition in the relevant literature: \emph{(1) manual decomposition} and \emph{(2) dataflow analysis}. We directly enhance the state-of-the-art by proposing a new mechanism to deduce components from existing software in a semiautomatic way. Security and soundness properties are composable and can be combined more easily than with any of the previously proposed methods.

\emph{(1) Manual decomposition} of cryptographic protocols into trusted and untrusted
components running on a microkernel have been done for IPsec~\cite{helmuth2005mikro} and IKE~\cite{schulz2009secure,burki2013ikev2}, which
required a lot of manual work and expert knowledge. Even though the results
likely improve security, no guarantees exist that the partitioning was done correctly.
For every new protocol or major software update, the partitioning
must be redone.

\emph{(2) Dataflow analysis} has been proposed to eliminate most of the manual effort. Several tools were developed to assist and partially
automate program partitioning. \emph{PrivTrans}~\cite{brumley2004privtrans}
performs static analysis of source code to partition programs into core
functionality and a monitor. \emph{ProgramCutter}~\cite{wu2013automatically} automatically
splits a program based on traces collected for multiple program runs. 
The authors of~\cite{rubinov2016automated} performed taint analysis on Android applications to
move critical portions into an isolated trusted execution environment. None
of the tools exploit specifics of cryptographic operations. This results in an unnecessarily
large TCB or requires unsafe manual removal of taints using expert knowledge.

\paragraph{\textbf{Isolation mechanisms}}

Execution of components has strong isolation requirements, which can be met using different techniques. Hardware extensions have been
introduced for isolating memory accesses~\cite{woodruff2014cheri,oleksenko2017intel}, by creating trusted execution
environments~\cite{costanintel,trustzone} and encrypting memory~\cite{kaplan2016amd}. Microkernels~\cite{klein2009sel4,fiasco}, micro
hypervisors~\cite{steinberg2010nova} and separation kernels~\cite{buerki2013muen} realize isolation by means of a trusted software layer,
monolithic operating systems have been extended to enhance process isolation~\cite{watson2010capsicum,smalley2013security,linuxcontainers}. We assume
component isolation and controlled interaction, but do not rely on any particular technology.
These mechanisms can be used as building blocks for fulfilling our requirements.

Restricting privileges of software regarding their ability to perform system calls as can be found in the principle of least privileges literature is \emph{not applicable}~\cite{Goldberg1996polp}. These results always assume a monolithic system that inherently comes with large trusted computing bases, contradicting our goal of having a small TCB that can be checked for correctness.

\section{\label{sec:design}Design}

In this section we introduce our methodology to partition a given software into components in a structured way. It is a combination of manual decomposition and dataflow analysis based on predicate logic. We naturally combine the benefits of both methods, namely we inherit the specificity of manual decomposition and the efficiency of dataflow analysis. However, instead of analyzing the software implementation directly, we create an annotated data flow representation termed {\em model} as basis for all further analysis. As a last step we group together parts of the model with similar properties into partitions.

Accordingly, there are four main steps involved to apply our methodology.
\begin{enumerate}[label=\textit{\Alph*.}]
  \item Generate an abstract {\bf\em{}model} of the software
  \item Annotate parts of it with {\bf\em{}predicates}
  \item Perform a constraint {\bf\em{}dataflow analysis}
  \item Run an {\bf\em{}adaptive partitioning} algorithm
\end{enumerate}

The following sections introduce and describe the involved steps in more detail.

\subsection{\label{subsec:outline}Model}

A software implementation of a security protocol inherently requires guarantees from its environment.
However, as stated in the previous sections, not all parts of the software require the same guarantees.
To calculate the actually required guarantees for any part of the software we model it as
a graph of interconnected \emph{primitives}. 
All primitives have a set of input and output ports with associated guarantees. These guarantees are
represented using variables in a predicate.

Some primitives might have different guarantee requirements
on its input and output ports. For example, a symmetric encryption algorithm could have the requirement to guarantee
confidentiality on its input ports (\emph{data} and \emph{key}), while not requiring confidentiality on its output.
To capture such a relation, predicates are assigned to primitives describing the relation of guarantees between
ports of a primitive and security goals.

Initially, only a few model variables are assigned manually to capture the assumptions of the environment.
To derive the required guarantees for all parts of the model, a system of equations consisting of all predicates needs to be solved.
If a solution is found, then the assigned variable values represent the necessary guarantees of the protocol.

As a next step, primitives
that require similar guarantees are identified. The identified primitives
may then be realized as an isolated component with well-defined interfaces.

\subsubsection{Primitives and channels}

A \emph{primitive} is a finite state machine reacting to stimuli from the
environment or from other primitives. Stimuli are asynchronous messages
received through the \emph{input ports} of a primitive. It may send messages
through its \emph{output ports}.  Primitives are loosely coupled and make no
assumptions about connected primitives.

The \emph{Env} primitive is special, as it denotes the boundary of our model.
Messages received by this primitive from the environment are forwarded to its
output port.  Messages sent to its input port are forwarded to the environment,
respectively. A networking component with an integrated TCP/IP stack can for
example be represented as an \emph{Env} primitive.

\begin{figure}[ht]
    \sf\tiny
    \centering
    \makebox[\linewidth][c]{
     \resizebox{0.9\linewidth}{!}{%
        \input{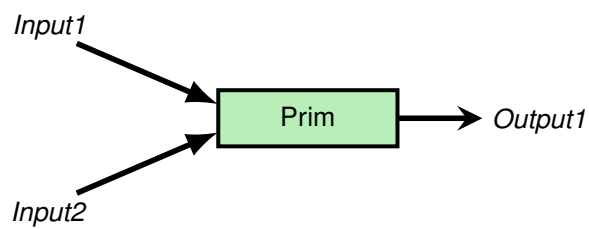}
     }
    }
    \caption{Example: Instance of primitive \emph{Prim} with two input ports
            \emph{Input1} and \emph{Input2} and one output port \emph{Output}.}
    \label{fig:primitive}
\end{figure}

Primitives are templates denoting a type of a node within the protocol model.
They carry specific semantics, input ports and output ports. To realize a protocol,
primitives are instantiated and their ports are connected to other instances
through \emph{channels} to form a graph. While \emph{instances} are unique,
the same primitive might be instantiated to multiple instances.
A \emph{channel} is a connection between exactly one output port of an instance to exactly one
input port of another instance. 


\begin{figure}
    \centering
    \tiny\sf
    \makebox[\linewidth][c]{
      \resizebox{1.0\linewidth}{!}{%
        \input{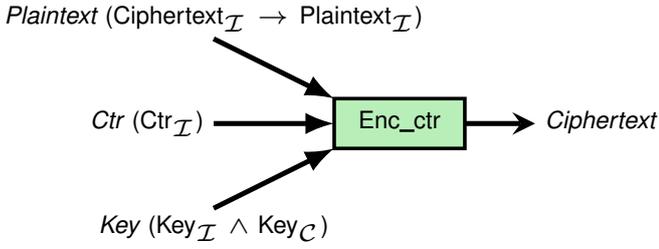}
      }
    }
    \caption{Counter-mode encryption primitive with predicates determining port guarantees}
    \label{fig:enc_ctr}
\end{figure}

\begin{figure}
    \begin{align}
    \R_{\text{enc\_ctr}} &= \encryptrule
    \end{align}
    \caption{Predicate determining guarantees for the counter-mode encryption primitive}
    \label{eq:encrypt}
\end{figure}

\subsection{Predicates}

\begin{figure}[ht!]
    \centering
    \tiny\sf
    \newcommand{\us}{Unserialize}
    \newcommand{\m}{\mod{}}
    \newcommand{\gr}{\gamma r}
    \newcommand{\gi}{\gamma i}
    \makebox[\linewidth][c]{
      \resizebox{0.96\linewidth}{!}{%
        \input{images/example_dh.teximg}
      }
    }
    \caption{DH key exchange before partitioning. \emph{Unknown}: white, \emph{No guarantees}: gray/continuous, \emph{Confidentiality}: red/dashed, \emph{Integrity}: blue/dotted, \emph{Both}: purple/dot-dashed}
    \label{fig:dhkex}
\end{figure}

\begin{figure}[ht!]
    \centering
    \tiny\sf
    \newcommand{\us}{Unserialize}
    \newcommand{\m}{\mod{}}
    \newcommand{\g}{\gamma}
    \newcommand{\gr}{\gamma r}
    \newcommand{\gi}{\gamma i}
    \makebox[\linewidth][c]{
      \resizebox{1.0\linewidth}{!}{%
        \input{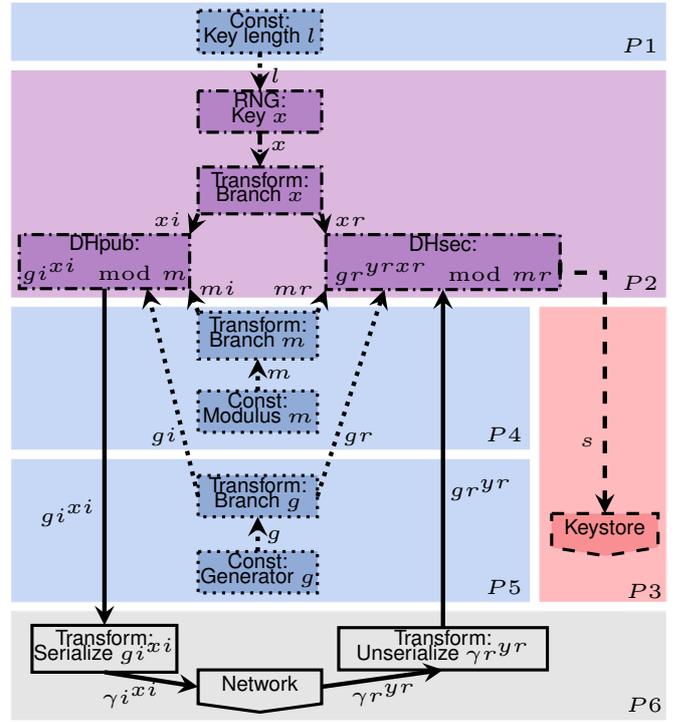}
      }
    }
    \caption{DH key exchange after partitioning. \emph{Unknown}: white, \emph{No guarantees}: gray/continuous, \emph{Confidentiality}: red/dashed, \emph{Integrity}: blue/dotted, \emph{Both}: purple/dot-dashed}
    \label{fig:dhkex_visual}

\end{figure}

An instance of a symmetric decryption primitive transforms ciphertext into
plaintext using a secret key. Intuitively, no confidentiality needs to be
guaranteed for the original ciphertext when sent to the decryption primitive,
as confidentiality is achieved by means of cryptography. But once the data has
been decrypted and is sent to an output port, confidentiality is no longer
achieved by cryptography, but must be guaranteed by the channel connected to
the output port and by the receiving instance.


For a hypothetical primitive $\PR{}_{Prim}$~(\Cref{fig:primitive}) with input ports $\invar{Input1}$
and $\invar{Input2}$ and an output port $\outvar{Output}$, a predicate 
$\R$, representing required guarantees, is expressed as $\R{}_{\text{Prim}} = \intg{\invar{Input1}} \land{}
\conf{\invar{Input2}} \land{} \intg{\outvar{Output}}$. This states, that the
primitive \emph{Prim} assumes that the input channel connected to $\invar{Input1}$
guarantees integrity (denoted by the $\intgsym$ in the subscript). For the respective channel of $\invar{Input2}$,
confidentiality ($\confsym$) is assumed. Furthermore, it assumes that the
channel associated with the output port $\outvar{Output}$ (marked
by an arrow symbol) 
guarantees integrity. 


Channels must provide at least the guarantees required by the ports they
connect. This implies that for two instances \emph{A} and \emph{B} where A's
output port $\outvar{Out}$ is connected to B's input port $\invar{In}$ through
a channel \emph{Chan} the predicate $\R{}_{\text{Chan}} = \intg{\outvar{\text{Out}}}
\Leftrightarrow \intg{\invar{\text{In}}} \land{} \conf{\outvar{\text{Out}}}
\Leftrightarrow \conf{\invar{\text{In}}}$ holds. For ease of understanding,
channel predicates are not given explicitly, but represented by identical port names
of connected ports throughout the remaining document.

\subsection{\label{subsec:examples}Dataflow analysis}

Using the introduced building blocks, a simple example for counter-mode
encryption is evaluated as well as a text-book Diffie-Hellman key agreement protocol.

\subsubsection{\label{subsubsec:simple_enc}Simple encryption}

The counter mode encryption primitive \emph{Enc\_ctr} in \Cref{eq:encrypt} has three input ports:
$\invar{Plaintext}$, $\invar{Key}$ and $\invar{Ctr}$ for receiving plaintext,
key and an initial counter, respectively. The result of the encryption is sent to the
$\outvar{Ciphertext}$ port.

The guarantees it requires from its runtime environment to fulfill its security
objective are defined in \Cref{eq:encrypt}: \eqref{eq:encrypt_1} If
integrity is required for $\outvar{Ciphertext}$ then integrity is also required
for the plaintext input port, as an attacker controlling the plaintext can
influence the ciphertext. Counter mode encryption does not achieve integrity
cryptographically, hence, integrity for the ciphertext is only achieved by
guaranteeing integrity for the plaintext. \eqref{eq:encrypt_2} Integrity and
confidentiality of the encryption key must always be guaranteed or an attacker
could learn or choose the key. \eqref{eq:encrypt_3} The integrity of the
counter is essential to counter mode, as using the same key/counter combination
twice is fatal.

\subsubsection{\label{subsubsec:simple_dh}Diffie-Hellman}

We use the Diffie-Hellman key agreement as a more sophisticated example
(\Cref{fig:dhkex}). Two connections to the environment exist in this model:
\emph{Keystore} for holding keys and \emph{Network} for sending and receiving messages over
an untrusted channel. As outlined in \Cref{subsec:outline} we formalize the
assumptions about the environment in \eqref{eq:gammax} and \eqref{eq:gammay}.
For brevity, we leave out channel predicates and relieve the \emph{Keystore} from the necessary integrity assumption.

\begin{figure}[ht]
    \scriptsize
    \begin{align}
        \conf{\gamma{}i^{xi}} \equiv \false&\land{}\intg{\gamma{}i^{xi}} \equiv \false\label{eq:gammax} \\
        \conf{\gamma{}r^{yr}} \equiv \false&\land{}\intg{\gamma{}r^{yr}} \equiv \false\label{eq:gammay} \\
        \intg{gi} \land{}& \intg{mi}  \land{} \intg{gr} \land{} \intg{mr} \label{eq:g_params} \\
        \conf{xr} \land{}& \intg{xr} \land{} \conf{xi} \land{} \intg{xi} \land{} \conf{s} \label{eq:g_keys} \\
        \conf{x} \land{}& \intg{l} \label{eq:rng} \\
        \intg{s} \rightarrow (\intg{g^y} \land{} \intg{gr} &\land{} \intg{mr} \land{} \intg{xr} \label{eq:dhseci})
    \end{align}
    \caption{Assumptions for the environment and predicates for cryptographic elements}
\end{figure}

\begin{figure}[ht]
    \scriptsize
    \begin{align}
        \conf{m}                   &\rightarrow (\conf{mi} \land \conf{mr}) \label{eq:confm} \\
        \intg{mi} \lor \intg{mr}   &\rightarrow \intg{m} \label{eq:intgmir} \\
        \conf{g}                   &\rightarrow (\conf{gi} \land \conf{gr}) \label{eq:confg} \\
        \intg{gi} \lor \intg{gr}   &\rightarrow \intg{g} \label{eq:intggir} \\
        \intg{\gamma{}^x}          &\rightarrow \intg{g^x} \label{eq:intggammax} \\
        \conf{x}                   &\rightarrow (\conf{xi} \land \conf{xr}) \label{eq:confx} \\
        \intg{xi} \lor \intg{xr}   &\rightarrow \intg{x} \label{eq:intgxir} \\
        \intg{gr^{yr}}             &\rightarrow \intg{\gamma{}r^{yr}} \label{eq:intggy} \\
        \conf{\gamma{}r^{yr}}      &\rightarrow \conf{gr^{yr}} \label{eq:confgammay} \\
        \intg{\gamma{}i^{xi}}      &\rightarrow \intg{gi^{xi}} \label{eq:intggammax} \\
        \conf{gi^{xi}}             &\rightarrow \conf{\gamma{}i^{xi}} \label{eq:confgx}
    \end{align}
    \caption{Predicates for Transform primitives used in Diffie-Hellman model}
    \label{eq:gxform}
\end{figure}


\begin{figure}[t]
    \scriptsize
    \begin{align}
        \intg{s}, \intg{g^y}&, \conf{l}, \conf{g}, \conf{m}, \conf{g^x} \gets \false \label{eq:as} \\
        \ref{eq:intggammax}, \ref{eq:gammax}:\              &\intg{g^x} \equiv \false \\
        \ref{eq:confgammay}, \ref{eq:gammay}:\              &\conf{g^y} \equiv \false \\
        \ref{eq:g_keys}, \ref{eq:rng}, \ref{eq:confx}:\     &\conf{xi}  \equiv \true, \intg{xi} \equiv \true \label{eq:xi} \\
        \ref{eq:g_params}:\                                 &\intg{mi}  \equiv \true, \intg{gi} \equiv \true \label{eq:as_gi} \\
        \ref{eq:g_params}:\                                 &\intg{mr}  \equiv \true, \intg{gr} \equiv \true \label{eq:as_gr} \\
        \ref{eq:g_keys}:\                                   &\conf{xr}  \equiv \true, \intg{xr} \equiv \true \label{eq:xr} \\
        \ref{eq:g_keys}:\                                   &\conf{s}   \equiv \true \\
        \ref{eq:intgmir}, \ref{eq:as_gi}, \ref{eq:as_gr}:\  &\intg{m}   \equiv \true \\
        \ref{eq:intggir}, \ref{eq:as_gi}, \ref{eq:as_gr}:\  &\intg{g}   \equiv \true \\
        \ref{eq:as}, \ref{eq:confm}:\                       &\conf{mi}  \equiv \false, \conf{mr} \equiv \false \\
        \ref{eq:as}, \ref{eq:confg}:\                       &\conf{gi}  \equiv \false, \conf{gr} \equiv \false \\
        \ref{eq:rng}:\                                      &\conf{x}   \equiv True \\
        \ref{eq:intgxir}, \ref{eq:xi}, \ref{eq:xr}:\        &\intg{x}   \equiv True
    \end{align}
    \caption{Solution for the above Diffie-Hellman model}
    \label{eq:dh_solution}
\end{figure}

The core of Diffie-Hellman is the \emph{DHpub} primitive calculating $g^x \mod
m$ and the \emph{DHsec} primitive calculating the shared secret $s = g^{yx} \mod
m$. The value for the secret key $x$ is created by a random number generator
\emph{RNG}. Fixed values like the length of $x$ or the values for $g$ and $m$
originate from \emph{Const} primitives. Encoding and branching of data to
several primitives is done by \emph{Transform}.

We can derive the complete guarantees for the model given in
\Cref{fig:dhkex} by using the guarantees of the environment and a couple of
generic primitive predicates. The \emph{DHpub} and \emph{DHsec} primitives require
integrity for their parameters and secret key, so that an attacker cannot
choose an own value (\ref{eq:g_params}). The secret key $x$ for both primitives must be
confidential, just like the resulting key $s$ (\ref{eq:g_keys}).

The result $x$ of $\mathit{RNG}$ must have confidentiality guaranteed and key
length $l$ is public, but it's integrity must be protected to prevent an
attacker from choosing a too short key (\ref{eq:rng}). We over-approximate the
relationship between inputs and outputs of $\emph{Transform}$: Whenever any
input requires confidentiality, all outputs require confidentiality. Whenever
an output requires integrity, all inputs must guarantee it. This implies the
relevant guarantees shown in \Cref{eq:gxform}.

To minimize the requirements for the implementation, we set variables lacking
explicit guarantees to $\false$ (\ref{eq:as}). Together with the predicates above,
we can determine a valid assignment for all variables in the model as shown in
\Cref{eq:dh_solution} and visualized in \Cref{fig:dhkex_visual}.

As stated above, our model does not assume integrity for the \emph{Keystore}.
This is a consequence of the Diffie-Hellman key agreement scheme, which does not achieve integrity by cryptographic means.
If we had chosen the \emph{Keystore} guarantees realistically, our method would have
yield a conflict between the absence of integrity stated for the network environment
and the system of equations given by our model. Specifically the assumption
$\intg{s}$ together with the predicates, \eqref{eq:dhseci}, \eqref{eq:intggy}, \eqref{eq:gammay}
result in the contradiction $\intg{\gamma{}r^{yr}} \land \neg{\intg{\gamma{}r^{yr}}}$.

\subsection{\label{subsec:extracting}Adaptive Partitioning}

The na\"{\i}ve transformation of our model into an implementation would create a single component for every instance
in the model. As communication between components is expensive, this results in a large communication overhead, compared to a monolithic implementation. 
To reduce this overhead we introduce an algorithm to trade-off communication cost against TCB size. Instances with compatible guarantees
can be merged into a single component.
Due to efficient communication within a component, the total overhead is reduced.
However, merging instances into fewer, but larger components comes at the cost of increasing the size of the TCB. 

We instantiate three simple partitioning algorithms to demonstrate the described trade-off capabilities.
One algorithm that only merges identical guarantees and two that take the semantics of the primitives into account.
By adopting an appropriate cost function, more powerful, adaptive partitioning schemes
can be realized.

\begin{description}

    \item[Merge Basic] The simplest approach is to put connected instances with
    identical guarantees into the same partition. The algorithm iterates over all
    instances in the model. For each instance, it checks whether it was assigned to
    a partition already. Otherwise, it assigns it to a new partition. For each new
    partition it traverses all connected components with the same guarantees and
    tries to assign them to the newly generated partition.

    \item[Merge Const] Additional to the basic partitioning scheme, this algorithm merges all
    \emph{Const} instances with their connected partition if that partition
    provides sufficient guarantees. This eliminates single instances of the \emph{Const} primitive
    representing an own partition as seen in $P1$ of \Cref{fig:dhkex_visual}. As constants
    are simple, merging them results only in a minor increase of the TCB.

  \item[Merge Branch] The third algorithm is similar to {\em Merge Const} in that it joins
    \emph{Const} primitives, as well as additional, simple instances into the same partition.
    To define \emph{simple} instances an arbitrary metric can be used, for examples the lines
    of code to implement an instance. If the metric used relates to the TCB size, it can be argued
    that the TCB is not increased significantly. As an example refer to the merging of partitions
    $P4$ and $P5$ into $P2$ in \Cref{fig:dhkex_visual}.
\end{description}

\section{\label{sec:scaling}Application of our framework}

\begin{figure*}[h!]
    \centering
    \tiny\sf
      \makebox[\linewidth][c]{
        \resizebox{1.0\linewidth}{!}{%
          \input{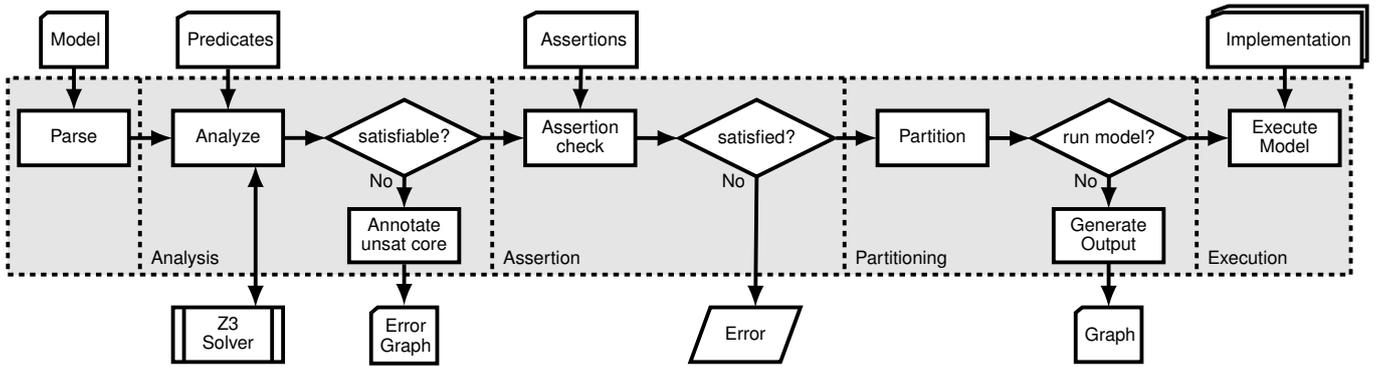}
        }
      }
    \caption{The four phases of the \projecttitle{} toolset}
    \label{fig:toolset}
\end{figure*}

As seen in the previous Diffie-Hellman example, the number of predicates to be
checked for a model quickly grows to an unmanageable size. In the following
section we describe our automatic \projecttitle{} toolset for partitioning and
asserting of protocol models and apply it to the OTR protocol. It consists of
four main phases: \emph{Analysis}, \emph{Assertion}, \emph{Partitioning}, and
\emph{Execution} (cf. \Cref{fig:toolset}). The last phase of the
\projecttitle{} toolset \emph{Execution} is not part of our methodical
framework as described in \Cref{sec:design}, but included to perform functional
validation.

\subsection{\projecttitle{} Toolset}

In the \emph{analysis phase}, model constraints are derived from primitive predicates,
channel predicates and the guarantees assumed for the environment. The
resulting constraint set is passed to an SMT solver, \textit{e.g.} Z3~\cite{de2008z3}, to find
a valid assignment for all guarantees. It may not be satisfiable, in which case
the solver may produce an unsatisfiability core. If available, this minimum set of constraints that lead to the
contradiction is mapped to the graph by our tool. The conflict can then be
inspected visually and used to find conflicts in the model or the assumptions.

The \emph{assertion phase} checks assertions associated with the model based
on independent expert knowledge. Guarantees are derived automatically in
the partitioning phase, solely using primitive predicates, channel predicates and
assumption about the environment. However, missing guarantees may render a
previously correct algorithm insecure. As assertions are not used in
partitioning, they serve as an independent sanity check for predicates and assumptions.

In the \emph{partitioning phase}, partitions are derived using the basic
recursive algorithm described in \Cref{subsec:extracting}. Optionally,
constants and branches are merged to reduce IPC overhead.

Lastly, models may be executed out of the \projecttitle{} toolset directly in the
\emph{execution phase}. We provide a library of primitives for
HMAC, DSA, Diffie-Hellman and more. Connections to the environment are realized
using TCP/IP, files or the console. Own implementations can be added easily by
implementing a Python class. Being solely a validation tool, all primitives
execute in the same Python process with no isolation enforced.

\subsection{Study: Off-the-record messaging}

We applied our approach to the Off-the-Record Messaging (OTR) protocol
\cite{borisov2004off} for secure instant messaging. As the Socialist
Millionaire Protocol is optional for the security of OTR, we leave this part
out.

From the protocol specification we derived a component model. OTR's goal to
serve as a plugin-solution for existing messaging protocols, turned out to be
helpful identifying the \emph{Env} primitives representing the boundaries
of the model: \emph{User Data}, \emph{Keystore} and \emph{Fingerprint}. They
all require confidentiality and integrity guarantees as they handle user
message, long-term keys or the identity of the remote party. The \emph{Network}
environment used to interface OTR with the Internet has no guarantees
whatsoever.

\begin{figure}[ht]
  \centering
  \tiny\sf
  \makebox[\linewidth][c]{
    \resizebox{\linewidth}{!}{%
      \input{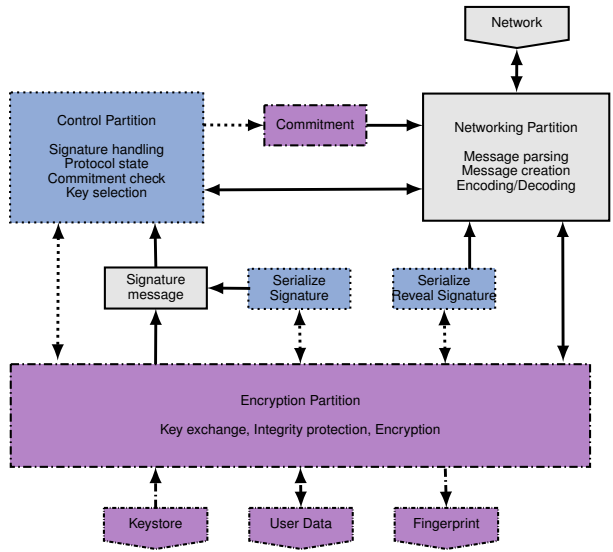}
    }
  }
  \caption{Partitions automatically derived from the OTR model (\emph{Integrity}: blue/dotted, \emph{Confidentiality}: red/dashed, \emph{Both}: purple/dot-dashed, \emph{No guarantees}: gray/continuous)}
  \label{fig:otr_partitions}
\end{figure}

\begin{figure}[ht]
  \centering
  \includegraphics[width=\linewidth]{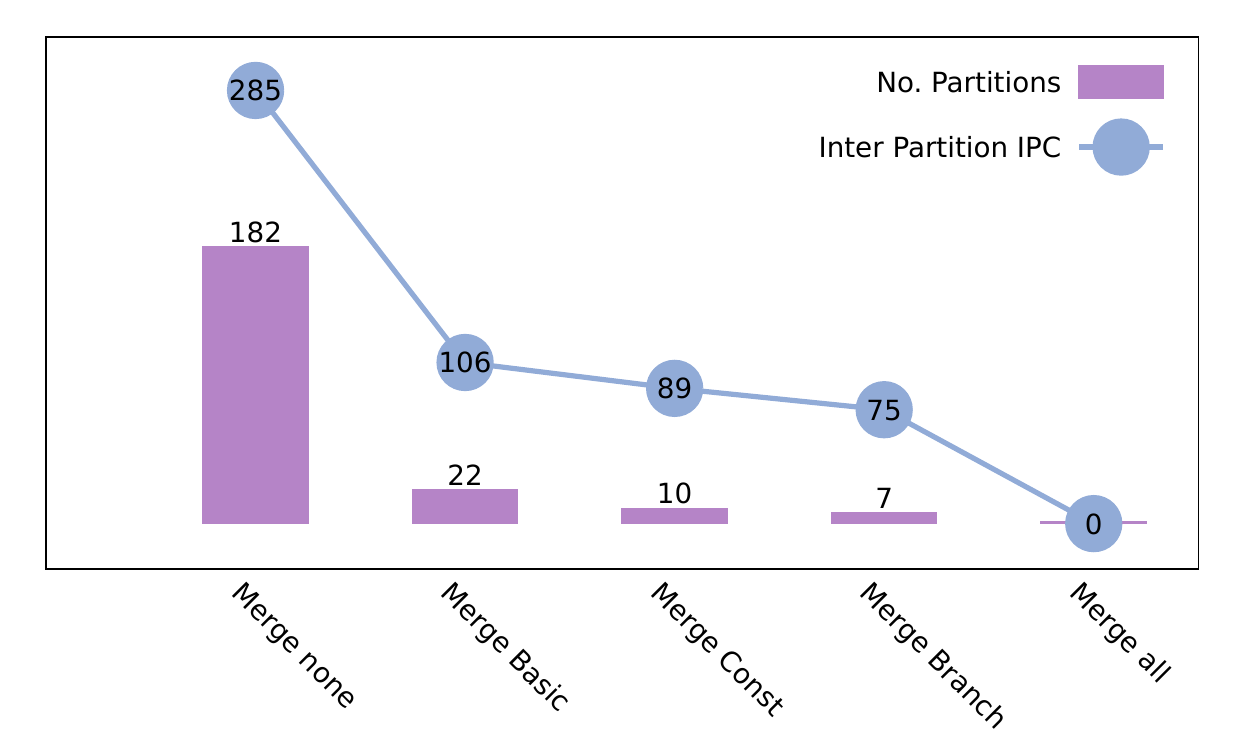}
  \caption{Process and IPC overhead in OTR}
  \label{fig:overhead}
\end{figure}

The model for the full OTR protocol consists of 186 primitive instances
connected by 285 channels. Only the four instances denoting the environment of
the model are annotated explicitly with guarantees. Based on that information
our optimized \emph{Merge Branch} algorithm (cf. \Cref{subsec:extracting})
automatically splits the protocol into the 7 partitions shown in
\Cref{fig:otr_partitions}.

In a step independent of the model creation, we revisited the OTR specification
and augmented the model by 46 assertions which are validated correct by the
\projecttitle{} toolset. The partitioned model runs against a minimal test client built with
the OTR reference implementation of the protocol authors~\cite{libotr}. We
successfully negotiate keys and send messages between both implementations.

\section{\label{sec:discussion}Discussion}

To assess the quality of our approach, we need to consider the overhead
caused by additional inter-process communication, the partitioning
introduced and the security advantage of partitioning a protocol.

\subsection{Overhead}

In a partitioned protocol, security is improved by isolating code with
different security guarantees.
That security gain comes at an added cost of inter-process
communication.
Our baseline is a monolithic implementation with one
single process and therefore efficient communication.

As shown in \Cref{fig:overhead}, simply implementing every instance within its own
component (\emph{Merge none}), expectedly leads to a prohibitively
large number of processes and inter-process communication for our OTR model\footnote{In
total, 182 processes with 285 inter-process channels would be needed to
implement the protocol.}.
To reduce the overhead we partition the model such that primitives with
the same guarantees are merged into a single component. That basic
partitioning schema \emph{Merge Basic} reduces the required number of
processes by 87\% and the amount of inter-process channels by 62\%.
Performance optimizations can be applied to further reduce the overhead.
As such, very small primitives like constants can be merged into
adjacent components using \emph{Merge Const}. This further reduces
processes and communication channels by 54\% and 16\%, respectively.
Similarly, \emph{Merge Branch} merges even more primitives.
This leads to the architecture depicted in
\Cref{fig:otr_partitions} with the minimum number of processes possible
without significantly increasing the TCB. A description of the partitioning schemes can be found in \Cref{subsec:extracting}.

Partitioning software may significantly increases the total number of processes,
thus consuming additional system resources. A large number of processes
is for example used to improve fault-isolation in web browsers, with acceptable
overhead~\cite{reis2008multi}. Further overhead is imposed by additional
inter-process communication, which is performed very efficiently on modern microkernels
\cite{steinberg2010nova}, typically orders of magnitude faster than common
 cryptographic primitives.

 \begin{table}
  \centering
  \setlength\tabcolsep{.5em}
    \begin{tabular}{@{}llll@{}} 
    \toprule
    \textbf{Guarantees}         & \textbf{Monolithic}   & \textbf{Partitioned}    & \textbf{Reduction} \\
    \hline
    None                        & -                     &  [7266]                 & - \\
    Integrity                   & -                     &  2645                   & - \\
    Confidentiality + Integrity & 10957                 &  3155                   & 29\% \\
    \hline
    \textbf{TCB}                & \textbf{10957}        &  \textbf{5800}          & \textbf{53\%} \\
    \bottomrule
    \end{tabular}
    \vspace{.5em}
    \caption{Complexity of partitioned vs. monolithic OTR.}
\end{table}



\begin{figure*}[ht!]
    \centering
    \tiny\sf
      \makebox[\linewidth][c]{
        \resizebox{1.0\linewidth}{!}{%
          \input{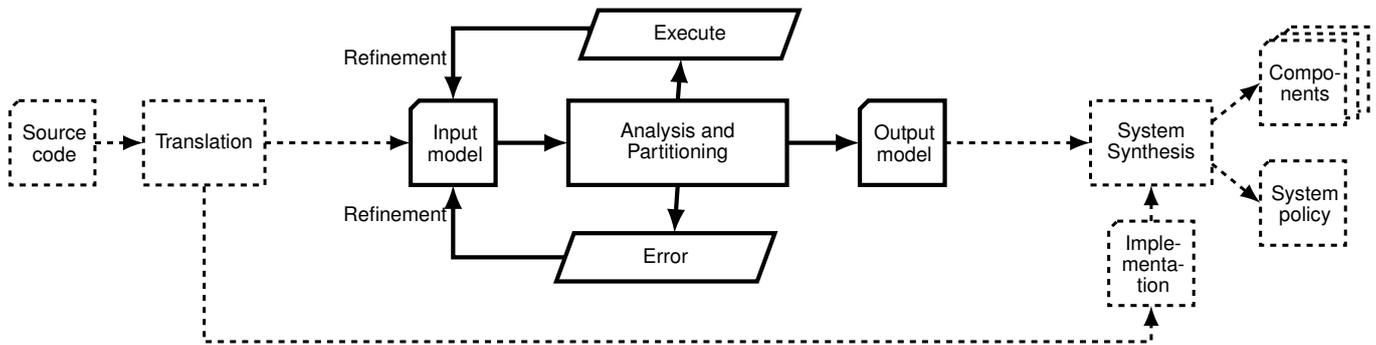}
        }
      }    
        
    \caption{High-level workflow for creating partitioned cryptographic protocols.
             Dashed parts mark future work items.}
    \label{fig:workflow}
\end{figure*}

\subsection{Security gain}

The security gain achieved by partitioning a security protocol can be measured
by the size of the resulting trusted computing base. Monolithic software
must guarantee confidentiality and integrity for their whole implementation, as
an error in a part unrelated to the security protocol can completely defeat security.

Our approach reduces the 
trusted computing base by assigning parts of the protocol to components without
security guarantees. Those parts become untrusted and can be implemented in the most
convenient way. Additionally, we treat confidentiality and integrity independently,
which results in components where only integrity needs to be guaranteed. While this
does not change the size of the overall TCB, it does in fact simplify assumptions
for the respective components.

In the lack of a full component-based implementation, we estimate the TCB size
using the code in our model execution framework. As it implements primitives as
Python classes, we chose a python OTR implementation to compare
against.\footnote{\url{https://github.com/python-otr/pure-python-otr}}
The TCB calculation includes cryptographic and standard python classes used by
the implementations, but excludes the python runtime an everything beneath.
TCB size is represented by measuring the source lines of code (SLOC\footnote{generated
using David A.  Wheeler's 'SLOCCount'}).

By partitioning the OTR protocol, we reduce the size of the OTR-related
TCB to 53\% compared to a monolithic Python implementation
using the same cryptographic libraries. Our approach of treating
confidentiality and integrity independently, specifically reduced the part of
the TCB for which confidentiality is to be guaranteed to only 29\% of the
original size.

\subsection{\label{subsec:future}Future work}

We automatically partition a security protocol, determine the guarantees
required for the partitions and derive a communication policy between them. The
user of our \emph{PrettyCat} toolset still needs to provide the input model, which involves
manual work and presents an additional source of errors. Also, while the
result of our analysis is a machine-readable, partitioned model, the
transformation into a component-based system is left as a manual step.

As suggested in \Cref{fig:workflow}, we are going to extend our tools to extract
the input model from existing source code using data flow analysis, e.g. FlowDroid
\cite{arzt2014flowdroid}. Preliminary results suggest that most
primitives can be matched against the source code automatically by annotating
cryptographic interfaces.

Additionally, we are implementing automatic synthesis of partitioned security
protocols for different target platforms, e.g. the Genode OS framework~\cite{feske2016genode}, including the
system structure, communication policy and instantiated primitives.
Even though implementations for common primitives will be provided, reusing
source code from model extraction during system synthesis will be investigated further.
The proposed extensions will reduce user interaction to manually annotating \emph{Env} primitives.

A mostly automatic partitioning process opens an opportunity to study other and
more complex security protocols. The widespread TLS protocol is an obvious and
worthwhile candidate, as are modern widely adopted secure messaging protocols
like Signal~\cite{perrin2016the} or OMEMO~\cite{verschoor2016omemo}.

\section{\label{sec:summary}Summary \& Conclusion}

Partitioning security protocols into component-based implementations with a
minimal TCB facilitates code review and verification. Traditionally this has
been done in a manual ad-hoc manner which is costly and error-prone. Existing
software partitioning tools are too imprecise, yield a large TCB and often
require unsafe manual steps.

We present a systematic approach to automatically partition security protocols.
Our method models them as primitives connected by channels and derives
required security guarantees using a constraint solver. A number of performance
optimizations help to trade-off between IPC overhead and TCB size without
sacrificing soundness.

We conclude that a methodology for automatically partitioning cryptographic
protocols into component-based systems is an important step towards trustworthy
systems. Our results indicate that an automation is desirable and feasible for
real world protocols.


\bibliographystyle{plain}
\bibliography{literature}

\begin{thebibliography}{10}

\bibitem{linuxcontainers}
{Linux Containers, accessed Feb. 14th 2017}.
\newblock \url{https://linuxcontainers.org/}.

\bibitem{libotr}
{OTR library and toolkit}.
\newblock \url{https://otr.cypherpunks.ca/}.

\bibitem{twitter2012securing}
{Securing your Twitter experience with HTTPS}.
\newblock
  \url{https://blog.twitter.com/2012/securing-your-twitter-experience-with-https},
  February 2012.

\bibitem{trustzone}
{ARM}.
\newblock {ARM TrustZone, accessed Feb. 14th 2017}.
\newblock \url{http://www.arm.com/products/security-on-arm/trustzone}.

\bibitem{arzt2014flowdroid}
Steven Arzt, Siegfried Rasthofer, Christian Fritz, Eric Bodden, Alexandre
  Bartel, Jacques Klein, Yves Le~Traon, Damien Octeau, and Patrick McDaniel.
\newblock {Flowdroid: Precise context, flow, field, object-sensitive and
  lifecycle-aware taint analysis for android apps}.
\newblock {\em Acm Sigplan Notices}, 49(6):259--269, 2014.

\bibitem{bhargavan2013implementing}
Karthikeyan Bhargavan, C{\'e}dric Fournet, Markulf Kohlweiss, Alfredo Pironti,
  and Pierre-Yves Strub.
\newblock {Implementing TLS with verified cryptographic security}.
\newblock In {\em {Security and Privacy (SP), 2013 IEEE Symposium on}}, pages
  445--459. IEEE, 2013.

\bibitem{borisov2004off}
Nikita Borisov, Ian Goldberg, and Eric Brewer.
\newblock {Off-the-record communication, or, why not to use PGP}.
\newblock In {\em Proceedings of the 2004 ACM workshop on Privacy in the
  electronic society}, pages 77--84. ACM, 2004.

\bibitem{brumley2004privtrans}
David Brumley and Dawn Song.
\newblock {Privtrans: Automatically partitioning programs for privilege
  separation}.
\newblock In {\em {USENIX Security Symposium}}, pages 57--72, 2004.

\bibitem{buerki2013muen}
Reto Buerki and Adrian-Ken Rueegsegger.
\newblock {Muen-an x86/64 separation kernel for high assurance}.
\newblock {\em University of Applied Sciences Rapperswil (HSR), Tech. Rep},
  2013.

\bibitem{burki2013ikev2}
Reto B{\"u}rki and Adrian-Ken R{\"u}egsegger.
\newblock {IKEv2 Separation: Extraction of security critical components into a
  Trusted Computing Base (TCB)}.
\newblock 2013.

\bibitem{cohn2016formal}
Katriel Cohn-Gordon, Cas Cremers, Benjamin Dowling, Luke Garratt, and Douglas
  Stebila.
\newblock A formal security analysis of the signal messaging protocol.
\newblock Technical report.

\bibitem{costanintel}
Victor Costan and Srinivas Devadas.
\newblock {Intel SGX explained}.
\newblock Technical report, Cryptology ePrint Archive, Report 2016/086, 2016.
  \url{https://eprint. iacr. org/2016/086}.

\bibitem{de2008z3}
Leonardo De~Moura and Nikolaj Bj{\o}rner.
\newblock {Z3: An efficient SMT solver}.
\newblock In {\em {International conference on Tools and Algorithms for the
  Construction and Analysis of Systems}}, pages 337--340. Springer, 2008.

\bibitem{devereaux2014data}
Ryan Devereaux, Glenn Greenwald, and Laura Poitras.
\newblock {Data pirates of the Caribbean: The NSA is recording every cell phone
  call in the Bahamas}.
\newblock {\em The Intercept}, 19, 2014.

\bibitem{fiasco}
TU~Dresden.
\newblock {The Fiasco microkernel, accessed Feb. 14th 2017}.
\newblock \url{https://os.inf.tu-dresden.de/fiasco/}.

\bibitem{feske2016genode}
Norman Feske.
\newblock {Genode Operating System Framework 16.05 -- Foundations}, 2016.

\bibitem{Goldberg1996polp}
Ian Goldberg, David Wagner, Randi Thomas, and Eric~A. Brewer.
\newblock A secure environment for untrusted helper applications confining the
  wily hacker.
\newblock In {\em Proceedings of the 6th Conference on USENIX Security
  Symposium, Focusing on Applications of Cryptography - Volume 6}, SSYM'96,
  pages 1--1, Berkeley, CA, USA, 1996. USENIX Association.

\bibitem{helmuth2005mikro}
Christian Helmuth, Alexander Warg, and Norman Feske.
\newblock {Mikro-SINA—Hands-on Experiences with the Nizza Security
  Architecture}.
\newblock {\em Proceedings of the DACH Security}, 2005.

\bibitem{kaplan2016amd}
David Kaplan, Jeremy Powell, and Tom Woller.
\newblock {AMD Memory Encryption Whitepaper}.
\newblock
  \url{http:///md-dev.wpengine.netdna-cdn.com/wordpress/media/2013/12/AMD_Memory_Encryption_Whitepaper_v7-Public.pdf},
  April 2016.

\bibitem{klein2009sel4}
Gerwin Klein, Kevin Elphinstone, Gernot Heiser, June Andronick, David Cock,
  Philip Derrin, Dhammika Elkaduwe, Kai Engelhardt, Rafal Kolanski, Michael
  Norrish, et~al.
\newblock {seL4: Formal verification of an OS kernel}.
\newblock In {\em {Proceedings of the ACM SIGOPS 22nd symposium on Operating
  systems principles}}, pages 207--220. ACM, 2009.

\bibitem{kobeissi2017automated}
Nadim Kobeissi, Karthikeyan Bhargavan, and Bruno Blanchet.
\newblock Automated verification for secure messaging protocols and their
  implementations: A symbolic and computational approach.
\newblock In {\em IEEE European Symposium on Security and Privacy (EuroS\&P).
  Available at http://prosecco. gforge. inria.
  fr/personal/bblanche/publications/KobeissiBhargavanBlanchetEuroSP17. pdf. To
  appear}, 2017.

\bibitem{KobeissiBhargavanBlanchetEuroSP17}
Nadim Kobeissi, Karthikeyan Bhargavan, and Bruno Blanchet.
\newblock Automated verification for secure messaging protocols and their
  implementations: A symbolic and computational approach.
\newblock In {\em 2nd IEEE European Symposium on Security and Privacy
  (EuroS\&P'17)}, Paris, France, April 2017. IEEE.
\newblock To appear.

\bibitem{koum2016end}
Jan Koum and Acton Brian.
\newblock {end-to-end encryption}.
\newblock \url{https://blog.whatsapp.com/10000618/end-to-end-encryption}, April
  2016.

\bibitem{meister2015how}
Andre Meister.
\newblock {How the German Foreign Intelligence Agency BND tapped the Internet
  Exchange Point DE-CIX in Frankfurt, since 2009}.
\newblock {\em netzpolitik.org}, March 2015.

\bibitem{oleksenko2017intel}
Oleksii Oleksenko, Dmitrii Kuvaiskii, Pramod Bhatotia, Pascal Felber, and
  Christof Fetzer.
\newblock {Intel MPX Explained: An Empirical Study of Intel MPX and
  Software-based Bounds Checking Approaches}.
\newblock {\em arXiv preprint arXiv:1702.00719}, 2017.

\bibitem{perrin2016the}
Trevor Perrin and Moxie Marlinspike.
\newblock {The Double Ratchet Algorithm}.
\newblock Technical report, November 2016.

\bibitem{reis2008multi}
Charlie Reis.
\newblock {Multi-process Architecture, Chromium Blog}.
\newblock Technical report, September 2008.

\bibitem{rubinov2016automated}
Konstantin Rubinov, Lucia Rosculete, Tulika Mitra, and Abhik Roychoudhury.
\newblock {Automated partitioning of android applications for trusted execution
  environments}.
\newblock In {\em {Proceedings of the 38th International Conference on Software
  Engineering}}, pages 923--934. ACM, 2016.

\bibitem{schillace2010default}
Sam Schillace.
\newblock {Default https access for Gmail, accessed Feb. 15th 2017}.
\newblock
  \url{https://gmail.googleblog.com/2010/01/default-https-access-for-gmail.html},
  January 2010.

\bibitem{schulz2009secure}
Steffen Schulz and Ahmad-Reza Sadeghi.
\newblock {Secure VPNs for trusted computing environments}.
\newblock In {\em {International Conference on Trusted Computing}}, pages
  197--216. Springer, 2009.

\bibitem{smalley2013security}
Stephen Smalley and Robert Craig.
\newblock {Security Enhanced (SE) Android: Bringing Flexible MAC to Android.}
\newblock In {\em {NDSS}}, volume 310, pages 20--38, 2013.

\bibitem{steinberg2010nova}
Udo Steinberg and Bernhard Kauer.
\newblock {NOVA: a microhypervisor-based secure virtualization architecture}.
\newblock In {\em {Proceedings of the 5th European conference on Computer
  systems}}, pages 209--222. ACM, 2010.

\bibitem{verschoor2016omemo}
Sebastian Verschoor.
\newblock {OMEMO: Cryptographic Analysis Report}.
\newblock Technical report, June 2016.

\bibitem{watson2010capsicum}
Robert~NM Watson, Jonathan Anderson, Ben Laurie, and Kris Kennaway.
\newblock {Capsicum: Practical Capabilities for UNIX.}

\bibitem{woodruff2014cheri}
Jonathan Woodruff, Robert~NM Watson, David Chisnall, Simon~W Moore, Jonathan
  Anderson, Brooks Davis, Ben Laurie, Peter~G Neumann, Robert Norton, and
  Michael Roe.
\newblock {The CHERI capability model: Revisiting RISC in an age of risk}.
\newblock In {\em {Computer Architecture (ISCA), 2014 ACM/IEEE 41st
  International Symposium on}}, pages 457--468. IEEE, 2014.

\bibitem{wu2013automatically}
Yongzheng Wu, Jun Sun, Yang Liu, and Jin~Song Dong.
\newblock {Automatically partition software into least privilege components
  using dynamic data dependency analysis}.
\newblock In {\em {Automated Software Engineering (ASE), 2013 IEEE/ACM 28th
  International Conference on}}, pages 323--333. IEEE, 2013.

\end{thebibliography}

\end{document}